\documentclass[twocolumn,prb,superscriptaddress,noshowpacs,epsf]{revtex4}

\usepackage{graphicx}

\begin{document}
\title{Hybridized solid-state qubit in the charge-flux regime}
\date{\today}
\author{J. Q. You}
\affiliation{Frontier Research System, The Institute of Physical
and Chemical Research (RIKEN), Wako-shi 351-0198, Japan}
\affiliation{Department of Physics and Surface 
Physics Laboratory (National Key Laboratory), 
Fudan University, Shanghai 200433, China}
\author{J. S. Tsai}
\affiliation{Frontier Research System, The Institute of Physical
and Chemical Research (RIKEN), Wako-shi 351-0198, Japan}
\affiliation{NEC Fundamental and Environmental Research Laboratories, Tsukuba,
Ibaraki 305-8051, Japan} 
%
\author{Franco Nori}
\affiliation{Frontier Research System, The Institute of Physical
and Chemical Research (RIKEN), Wako-shi 351-0198, Japan}
\affiliation{Center for Theoretical Physics, Physics Department, 
Center for the Study of Complex Systems, 
University of
Michigan, Ann Arbor, MI 48109-1140, USA} 

\begin{abstract}
Most superconducting qubits operate in a regime dominated by either 
the electrical charge or the magnetic flux.
Here we study an intermediate case: 
a hybridized charge-flux qubit with a third Josephson 
junction (JJ) added into the SQUID loop of the Cooper-pair box.
This additional JJ allows the optimal design of a low-decoherence qubit. 
Both charge and flux $1/f$ noises are considered. 
Moreover, we show that an efficient quantum measurement of either 
the current or the charge can be achieved by using different 
area sizes for the third JJ.             
\end{abstract}
\maketitle

\section{Introduction}

Solid-state qubits based on Josephson-junction (JJ) circuits have attracted 
considerable
attention in recent years and different kinds of Josephson qubits are being
explored by taking advantage of the charge and phase (flux) degrees of freedom.
Experimentally, quantum oscillations were observed in charge,~\cite{NAKA} 
phase,~\cite{HAN} and flux qubits.~\cite{CHIO}
A Josephson qubit in the intermediate regime between charge and flux 
also exhibited quantum oscillations~\cite{VION} 
and showed a high quality factor corresponding to a decoherence time 
of about $0.5$~$\mu$s.   
Because quantum information processing requires states to evolve coherently in 
a sufficiently long time, it is thus crucial to obtain qubits 
with very low decoherence.

Here we study a new type of Josephson qubit, somewhat similar to that 
in Ref.~\onlinecite{VION}, in which a third JJ is added into the SQUID 
loop of the Cooper-pair-box (CPB) qubit.~\cite{NAKA} 
In Ref.~\onlinecite{VION} this third JJ is connected to a current source 
and only used for measuring the quantum states of the CPB qubit.
Moreover, because this JJ is so large, the quantum states of the CPB qubit 
are only very slightly modified by it. 
Actually, without a bias current, the large third JJ can be 
approximated by a harmonic oscillator and the whole system can thus be considered
as a CPB qubit coupled to the oscillator. 
This is very similar to a CPB qubit in a cavity.~\cite{FN,Yale}
Very recently, the coherent dynamics of a flux qubit coupled to a harmonic oscillator 
has been studied,~\cite{FLUX} where the large-size SQUID connected to the qubit plays
the role of the harmonic oscillator.
 
In our present work, the CPB qubit is working in the charge-flux regime, 
as in Ref.~\onlinecite{VION}, but now the third JJ is {\it not} necessarily large 
and more importantly it is {\it not} used just as a measuring component. 
This additional degree of freedom in designing the charge-flux qubit 
allows us to optimize the qubit  
by changing the size of the third JJ. Indeed, here we show that 
the charge-flux qubit is gradually {\it hybridized} 
(in the quantum mechanical sense) with the third JJ 
when the area size of this additional JJ decreases. 
More importantly, we find that the qubit can be optimized to have  
the lowest decoherence at a suitable size of the third JJ.  
Furthermore, we show that efficient quantum measurements of either the 
current or the charge can be implemented by just choosing different sizes 
for the third JJ. 

The paper is organized as follows. In Sec.~II, we present the model Hamiltonian 
for the hybridized charge-flux qubit and study its properties. Section~III shows
the energy spectra of the qubit for different sizes of the third 
junction added into the SQUID loop of the CPB. 
To analyze the effects of different kinds of noises on the qubit, 
we employ the boson bath model in which a noise is described by a collection of 
spectrally distributed harmonic oscillators.  
The characteristic times for relaxation, decoherence, and leakage of the qubit 
states are calculated in Sec.~IV. 
We optimize the qubit to have the lowest decoherence by choosing a suitable size 
for the third junction. Section~V is devoted to quantum measurement. 
We propose two readout schemes to efficiently discriminate qubit states 
by taking advantage of the charge and flux degrees of freedom. 
Finally, conclusions are presented in Sec.~VI.

\section{The model}

The hybridized charge-flux qubit is shown in Fig.~1(a). 
The third JJ, i.e., the left one, is added into the SQUID loop 
of the CPB in which an island (denoted by a black dot) 
is connected by two JJ and coupled to a gate voltage by a capacitance $C_g$. 
When 
$C_1+C_g=C_2$,
the Hamiltonian of the system is given by
\begin{equation}
H=E_{cp}(N-n_g)^2+E_l(N_3+{1\over 2}n_g)^2+U,
\label{hamil}
\end{equation}
with 
\begin{equation}
U=\sum_{i=1}^3 E_{Ji}(1-\cos\phi_i).
\end{equation} 
Here  
\begin{eqnarray}
&&E_{cp}=2E_c,\;\;\;\; E_c={e^2\over 2C_2}; \nonumber\\
&&E_l={8 C_2E_c \over C_2+2C_3}. 
\end{eqnarray}
The phase drops through the three junctions are contrained by 
\begin{equation}
\phi_1-\phi_2+\phi_3+2\pi f_e=0, 
\end{equation}
where 
\begin{equation}
f_e={\Phi_e\over\Phi_0} 
\end{equation}
is the reduced magnetic flux
in the qubit loop (in units of the flux quantum $\Phi_0=h/2e)$.  
The operator 
\[
N=-i{\partial\over\partial\phi} \;,
\;\;\; \phi={1\over 2}(\phi_1+\phi_2),
\]
corresponds to the number of Cooper pairs on the island, and 
\[
N_3=-i{\partial\over\partial\phi_3}
\]
corresponds to the number of Cooper pairs tunneling through the left JJ.   
Here we consider the simpler case with 
$E_{J1}=E_{J2}=E_J$, $C_1=C_2=C$, $E_{J3}=\alpha E_J$, 
and $C_3=\beta C$. 
In this case, the periodic potential $U(\phi,\phi_3)$ is
\begin{equation}
U=E_J[(2+\alpha)-2\cos\phi\cos(\pi f_e+{1\over 2}\phi_3)-\alpha\cos\phi_3], 
\label{potent}
\end{equation}
and the condition 
$C_1+C_g=C_2$
can be approximately achieved because $C_g\ll C_1, C_2$.

\begin{figure}
\includegraphics[width=3.4in,
bbllx=76,bblly=208,bburx=545,bbury=705]{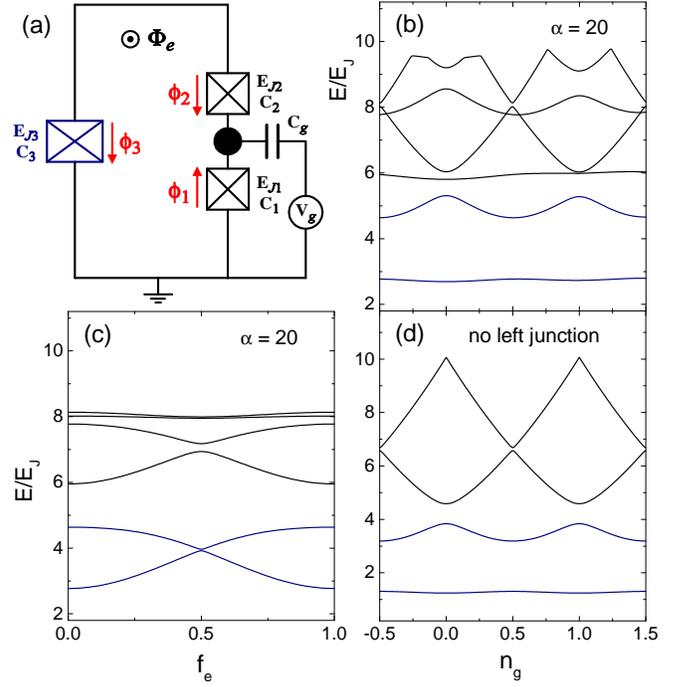}
\caption{(Color online) (a)~Schematic diagram of the hybridized charge-flux qubit, 
which consists of three 
JJs in a superconducting loop (pierced by an external magnetic flux $\Phi_e$) 
and a superconducting island (denoted by a black dot) coupled to a gate voltage 
via a capacitance $C_g$. 
The Josephson energies and capacitances of the JJs are $E_{J1}=E_{J2}=E_J$,
$C_1=C_2=C$, $E_{J3}=\alpha E_J$, and $C_3=\beta C$. 
Here we consider the charge-flux regime when $E_J=E_c\equiv e^2/2C_2$.
Unless explicitly stated otherwise, 
$\alpha=\beta$ is chosen throughout the paper. 
Energy levels of the charge-flux qubit versus (b) $n_g$ at $f_e=0$, 
and versus (c) $f_e$ at $n_g=0.5$, where $\alpha=\beta=20$. 
(d) Energy levels of the CPB qubit versus $n_g$ at $f_e=0$, 
without the left JJ. 
} 
\label{fig1}
\end{figure}

Assuming that the eigenstate of Hamiltonian (\ref{hamil}) has the form
as follows
\begin{equation}
\Psi(\phi,\phi_3)=e^{in_g(\phi-{1\over 2}\phi_3)}\psi(\phi,\phi_3),
\end{equation} 
one can cast the equation for eigenvalues and eigenfunctions  
\begin{equation}
H\Psi(\phi,\phi_3)=E\Psi(\phi,\phi_3) 
\end{equation}
to a standard Schr{\"o}dinger equation with a periodic potential:
\begin{equation}
H_0\psi(\phi,\phi_3)=E\psi(\phi,\phi_3), 
\label{eig}
\end{equation}
where
\begin{equation}
H_0=E_{cp}N^2+E_pN_3^2+U(\phi,\phi_3), 
\end{equation}
with $U(\phi,\phi_3)$ given by Eq.~(\ref{potent}).

Similar to a flux qubit (see, e.g., Refs.~\onlinecite{ORL} and \onlinecite{YOU}), 
the reduced Hamiltonian $H_0$ is just like that for a particle in a two-dimensional 
periodic potential, so the solution of Eq.~(\ref{eig}) has the Bloch-wave form
\begin{equation}
\psi(\phi,\phi_3)=e^{i(k\phi_p+k_3\phi_3)}u_{\bf K}(\phi,\phi_3),
\end{equation}
where ${\bf K}=(k,k_3)$. The constraint 
\begin{equation}
(k,k_3)=(-n_g,{1\over 2}n_g)
\end{equation}
on the wave vectors gives rise to  
\begin{equation}
\Psi(\phi,\phi_3)=u_{\bf K}(\phi,\phi_3),
\end{equation}
which ensures that $\Psi(\phi,\phi_3)$ is periodic in the phases 
$\phi$ and $\phi_3$.

Moreover, the Hamiltonian (\ref{hamil}) can be rewritten as 
\begin{equation}
H=H_{cp}+H_l+H_I, 
\end{equation}
where
\begin{equation}
H_{cp}=E_{cp}(N-n_g)^2+2E_J[1-\cos\phi\cos(\pi f_e)] 
\end{equation}
is the Hamiltonian of a CPB qubit,
i.e., the qubit with the left JJ absent in Fig.~1(a), and
\begin{equation} 
H_l=E_l(N_3+{1\over 2}n_g)^2+\alpha E_J(1-\cos\phi_3) 
\end{equation}
is the effective Hamiltonian of the left JJ. The interaction Hamiltonian 
\begin{equation}
H_I=2E_J\cos\phi\;[\cos(\pi f_e)-\cos(\pi f_e-{1\over 2}\phi_3)]
\end{equation} 
represents the coupling between the CPB qubit and the left JJ.
   
For a large left JJ, the phase drop $\phi_3$ is small, so the left JJ
can be approximated as a harmonic oscillator with frequency 
\begin{equation}
\Omega={4\over\hbar}(\kappa  E_JE_c)^{1/2}, 
\end{equation}
where 
\begin{equation}
\kappa  ={\alpha\over 1+\beta}.
\end{equation} 
Also, the interaction Hamiltonian can be approximated by 
\begin{equation}
H_I = -[\phi_3\sin(\pi f_e)-{1\over 4}\phi_3^2\cos(\pi f_e)]E_J\cos\phi,
\end{equation}
with 
\begin{equation}
\phi_3=\left[{4E_c\over\alpha (1+\beta)E_J}\right]^{1/4}(a+a^{\dag}),
\end{equation}
where $a$ ($a^{\dag}$) is the operator for annihilating (creating) a boson. 
Because $\alpha$ and $\beta$ are large for a large-area left JJ, 
it is clear that 
when $f_e\ne 0$, $H_I$ is dominated by a weak one-boson process, while 
a much weaker two-boson process is involved in $H_I$ for $f_e=0$.

\begin{figure}
\includegraphics[width=3.3in,
bbllx=84,bblly=228,bburx=498,bbury=707]{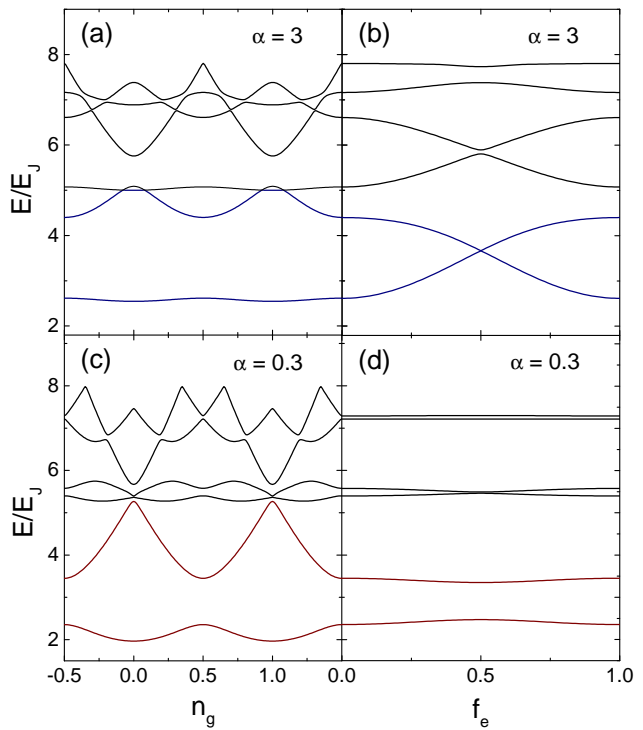}
\caption{(Color online) Energy levels of the charge-flux qubit versus $n_g$ 
at $f_e=0$ for (a) $\alpha=3$ and (c) 0.3, and 
versus $f_e$ at $n_g=0.5$ for (b) $\alpha=3$ and (d) 0.3.}
\label{fig2}
\end{figure}

\section{Energy spectrum} 

Below we show the hybridizing effects 
of the left JJ on the energy spectrum of the qubit 
in the charge-flux regime with $E_J=E_c$. 
The energy levels for $f_e=0$ and $n_g=0.5$ are given in Figs.~1(b) and 1(c),  
where a large left JJ with $\alpha=\beta=20$ is chosen. 
In contrast to the energy levels of the CPB qubit [cf. Fig.~1(d)], 
there exist additional levels due to the left JJ. 
However, because the left JJ is now large (i.e., $E_{J3}=20E_J$), 
the interaction between this JJ and the CPB qubit is small. 
Therefore, the energy levels of the CPB qubit are slightly 
modified by these additional levels, especially for the two lowest levels 
used for the qubit.

Figures~ 2(a) and 2(b) display the energy levels for $f_e=0$ and $n_g=0.5$ 
and a much smaller $E_{J3}$, since now $\alpha=3$.
The levels of the left JJ now hybridize with those 
above the two lowest levels, 
but the two lowest levels are still barely modified [comparing Fig.~2(a) with Fig.~1(d)].
This means that, as far as the two lowest states are concerned, the left JJ
with $\alpha=3$ can still be regarded as a large JJ. 
When the left JJ becomes even smaller (e.g., $\alpha=0.3$), $H_I$ 
becomes larger and the energy levels of both the left JJ and  
the CPB qubit become heavily hybridized [see Figs. 2(c) and 2(d)]; 
one can see that the energy levels in Fig.~2(c) look different 
from those in Fig.~1(b), but the two lowest levels can also be used for a qubit.

\section{State coherence and qubit optimization} 

Realistic qubit circuits
will experience fluctuations from both charge and magnetic flux. 
These noises will affect the coherence of the qubit states in the subspace 
with basis states $|0\rangle$ and $|1\rangle$, corresponding to the two lowest 
levels. To characterize the qubit-state coherence, 
the relaxation time $T_1$ and decoherence time $T_2$ are used:~\cite{BURK}
\begin{eqnarray}
&&{1\over T_1}\,\,=\,\,4|\langle 0|A|1\rangle|^2S(\omega_{01}), \nonumber\\
&&{1\over T_2}\,\,=\,\,{1\over 2T_1}+{1\over T_{\varphi}},
\label{deco}
\end{eqnarray}
with
\begin{equation}
{1\over T_{\varphi}}\,\,=\,\,|\langle 0|A|0\rangle-\langle 1|A|1\rangle|^2
S(\omega)|_{\omega\rightarrow 0}.
\end{equation}
Here $A$ is an operator characterizing the coupling between the qubit and 
the environment, and $S(\omega)$ is the power spectrum of the noise.
Moreover, because there are other levels above the lowest two, leakages from 
the qubit-state subspace to these outside levels can occur.
Therefore, two additional times:~\cite{BURK} 
\begin{eqnarray}
{1\over T_{Lk}}\!&\,=\,&\!4\sum_n|\langle n|A|k\rangle|^2S(\omega_{kn}), \nonumber\\
&& k=0,1, \;\;n=2,3,\cdots,
\end{eqnarray}
are needed to characterize the noise-induced transitions from the two lowest levels 
to the ones above. 

These results are based on the boson bath model in which the noise 
is described by a collection of harmonic oscillators with a spectral distribution. 
When Eq.~(\ref{deco}) is applied to a $1/f$ noise (see, e.g., Ref.~\onlinecite{BERT}),
the very low frequencies are cut off for the power spectrum $S(\omega)$.
This cutoff low-frequency part corresponds to the limit of  
very slow processes. For instance, for the $1/f$ charge noise, this can 
correspond to the extremely slow switchings of the trapped charges. 
If these fluctuating processes are much slower than the decoherence time $T_2$
of the qubit, 
they remain approximately static 
during the quantum operation and yield negligible dephasing.

\subsection{Johnson-Nyquist noises}

For Johnson-Nyquist noises, such as the fluctuations 
from gate voltage and external magnetic flux, 
the operators $A$ are given, respectively, by  
\begin{equation}
A_V={{E_{cp}N-E_lN_3}\over \sqrt{E_{cp}^2+E_l^2}},
\end{equation}
and
\begin{equation}
A_{\Phi}=\cos(\phi)\sin(\pi f_e+{1\over 2}\phi_3). 
\label{operator}
\end{equation}
The power spectrum is given by 
\begin{equation}
S(\omega)\equiv J(\omega)\coth\left({\hbar\omega\over 2k_BT}\right), 
\end{equation}
where $J(\omega)$ is the bath spectral density. 

For gate-voltage fluctuations characterized by an impedance $Z(\omega)$, 
the bath spectral density is
\begin{equation}
J_V(\omega)={2\pi\xi\over R_Q}\omega{\rm Re}[Z(\omega)],
\end{equation} 
where 
\begin{equation}
\xi=\left[1+{1\over(1+2\alpha)^2}\right]\left(C_g\over C_2\right)^2,
\end{equation} 
and 
$R_Q=h/e^2\approx 25.8$~k$\Omega$
is the quantum resistance. Here we choose $C_g=0.01C_2$, and consider the typical Ohmic 
case of $Z(\omega)=R_V=50$~$\Omega$. The external magnetic flux in the qubit loop 
is produced by a coil of inductance $L$ and resistance $R_L$. 
The bath spectral density of the external magnetic flux fluctuations is 
\begin{equation}
J_{\Phi}(\omega)={\pi \over 2}\left({R_Q\over R_L}\right)
{\eta^2\omega\over [1+(\omega L/R_L)^2]},
\end{equation}
where 
\begin{equation}
\eta={MI_c\over\Phi_0}, 
\end{equation}
with $I_c={2\pi E_J/\Phi_0}$, and $M$ is the mutual inductance 
between the qubit loop and the coil. Here we choose $E_J/h=20$ GHz, $R_L=100$ $\Omega$, 
$L=30 $ pH, and $M=5$ pH. These parameters correspond to realistic circuits.

\begin{figure}
\includegraphics[width=3.3in,
bbllx=56,bblly=179,bburx=500,bbury=700]{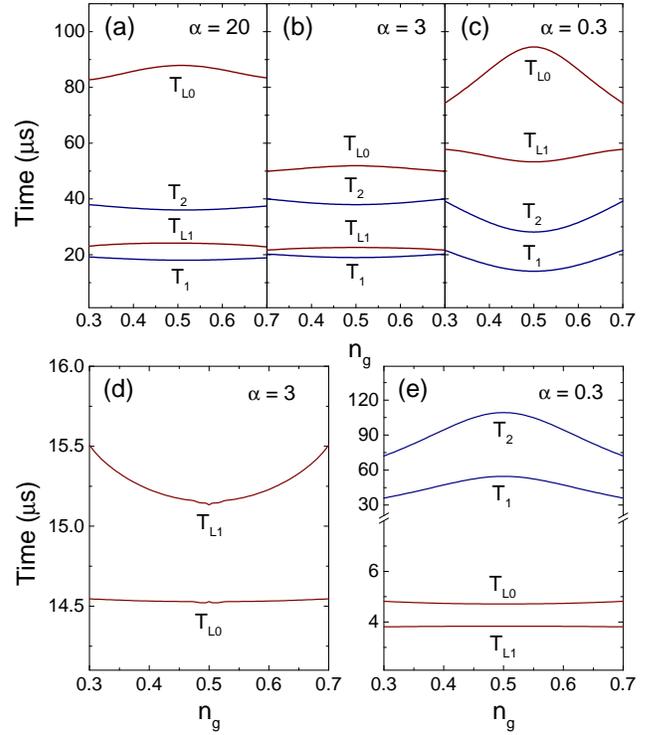}
\caption{(Color online) Relaxation ($T_1$), decoherence ($T_2$), 
and leakage ($T_{L0}$ and $T_{L1}$) times versus $n_g$ at (a) $\alpha=20$, 
(b) 3, and (c) 0.3 for gate-voltage noise, 
and at (d) $\alpha=3$ and (e) 0.3 for flux noise, where $f_e=0$.
In (d), $T_1$ and $T_2$ are not shown because they are 5 orders of 
magnitude larger than $T_{L0}$ and $T_{L1}$.
Here the temperature is chosen to be $T=30$~mK.}
\label{fig3}
\end{figure}

Figures 3(a)-3(c) show the four characteristic times at 
$\alpha=20$, 3, and 0.3 for the gate-voltage noise. These four times are almost 
of the same order of magnitude for different sizes of the left JJ;
especially $T_1$ (the minimum of them) and $T_2$ do not change much. 
This implies that the gate-voltage noise is mainly determined 
by the ratio of $E_J/E_c$ and less sensitive to the variation of the left JJ.  
The observation that $T_{L0}$ and $T_{L1}$ have almost the same order of magnitude 
as $T_1$ and $T_2$ also means that in this case the leakages produce equivalently 
important effects on the qubit states as the relaxation and decoherence in the 
qubit-state subspace. However, for the noise due to external flux fluctuations, 
the leakages dominate over the relaxation and decoherence [cf. Figs. 3(d) and 3(e)]. 
Moreover, when the external magnetic flux is around zero, 
the effects of the external flux noise are sensitive to the variation of the left JJ.
For instance, when the left JJ 
decreases in size to $\alpha=3$, the leakage times in the flux-noise case 
are comparable to the relaxation time $T_1$ in the case of gate-voltage noise  
[comparing Fig. 3(d) with Figs. 3(a)-3(c)], and the qubit-state leakages become 
more serious with $\alpha$ decreasing further [see Fig. 3(e)].

The above numerical results for the Johnson-Nyquist noises show 
that {\it the decoherence time} $T_2 \agt 30$~$\mu$s or more, 
{\it much longer than the experimental value} $T_2\approx 0.5$~$\mu$s in 
Ref.~\onlinecite{VION}.
This indicates that they could not be the major sources of decoherence 
in the charge-flux qubit. 
Instead, because the $1/f$ noise may be the main source of decoherence, 
we further study its effects on the charge-flux qubit.

\subsection{$1/f$ charge and flux noises}

There have been numerous attempts to model $1/f$ noise; including using a collection 
of independent bistable fluctuators with a given distribution of flipping 
rates~\cite{PAL,GAS} or by interacting two-level classical fluctuators.~\cite{NG} 
Alternatively, one can also model it using a boson bath with a $1/f$ 
spectral density.~\cite{MS} 
For the charge-flux qubit considered here, there can be two independent 
$1/f$ charge noises related with the background charge fluctuations of the CPB 
and the left JJ; 
the leakage rates $1/T_{Lk}$ as well as the relaxation and decoherence rates 
$1/T_i$ ($i=1,2$) are the sum of their respective contributions.
These two charge noises can be characterized by the power spectra 
\begin{eqnarray}
&& S_{q,cp}(\omega)=\left({2E_{cp}\over\hbar e}\right)^2
{\alpha_q\over\omega} \,, \nonumber\\ 
&& S_{q,l}(\omega)=\left({2E_l\over\hbar e}\right)^2
{\alpha_q\over\omega} \,, 
\end{eqnarray}
with the corresponding operators $A$ being  
\begin{eqnarray}
A_{q,cp} \!&=&\! -i{\partial\over\partial\phi} \,, \nonumber\\
A_{q,l} \!&=&\! -i{\partial\over\partial\phi_3} \,.
\end{eqnarray}
Here, for simplicity, $\alpha=\beta$, and $\alpha_q$ is chosen here 
to be the same for 
these two charge noises. In Ref.~\onlinecite{VION}, $\beta>\alpha$ because a current 
source is connected in paralell to the left JJ; this decreases 
$E_l\equiv 4E_{cp}/(1+\beta)$, and the dephasing due to the $1/f$ charge noise 
of the left JJ is weaker than that of $\beta=\alpha$.
Also, we can define a power spectrum for the $1/f$ flux noise:
\begin{equation}
S_{\Phi}(\omega)=\left({2\pi E_J\over\hbar\Phi_0}\right)^2
{\alpha_{\Phi}\over\omega} \,.
\end{equation}
The corresponding operator $A$ is 
\begin{equation}
A_{\Phi}=\cos(\phi)\sin(\pi f_e+{1\over 2}\phi_3) \,, 
\end{equation}
which is identical to Eq.~(\ref{operator}).

\begin{figure}
\includegraphics[width=3.2in,
bbllx=121,bblly=493,bburx=461,bbury=790]{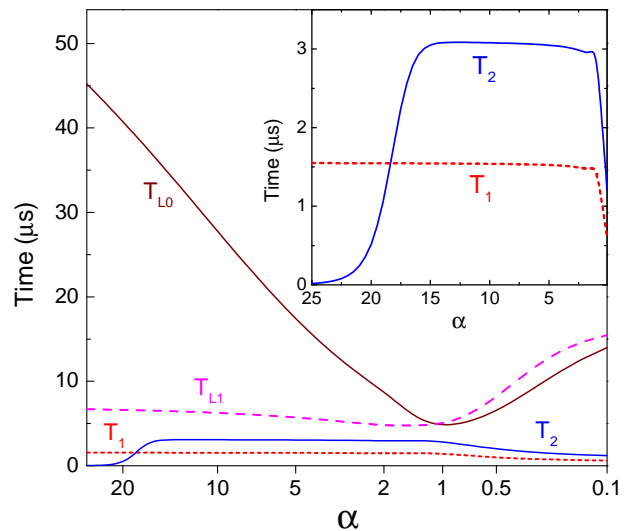}
\caption{(Color online) Relaxation ($T_1$), decoherence ($T_2$), 
and leakage ($T_{L0}$ and $T_{L1}$) times versus $\alpha$ at $(n_g,f_e)=(0.5,0)$ 
in the presence of $1/f$ charge noise, where $E_J/h=20$~GHz is chosen.
Inset:~$T_1$ and $T_2$ are replotted with $\alpha$ scaled linearly.}
\label{fig4}
\end{figure}

Figure~4 shows the four characteristic times  at the degeneracy point 
$(n_g,f_e)=(0.5,0)$ in the presence of the $1/f$ charge noise. We choose 
$\alpha_{q}=(0.7\times 10^{-3}e)^2$ for the power spectrum, which is very close to 
the value used for fitting the experimental data of 
the $1/f$ noise in the charge qubit.~\cite{NAKA1}
The cutoff frequency is chosen to be $\omega_c/2\pi=60$~Hz, 
corresponding to a time scale $\sim 2\times 10^4$~$\mu$s, much slower than 
the experimentally measured decoherence time $0.5$~$\mu$s of the charge-flux 
qubit.~\cite{VION}
To compare the effects of both charge and flux noises, 
we take the same cutoff frequency for the $1/f$ flux noise. 
Moreover, we use $\alpha_{\Phi}=3\times 10^{-12}\Phi_0^2$ for the flux-noise
power spectrum, which is the experimentally determined value of 
the flux qubit.~\cite{BERT}
In Fig.~4, the obtained leakage times $T_{L0}$ and $T_{L1}$ are longer than 
$T_1$ and $T_2$. This means that the leakage is not significant for the $1/f$ 
charge noise, even though the two lowest levels for the qubit are not very 
separated from the higher levels (cf.~Figs.1 and 2).

We also calculated the four characteristic times for the $1/f$ flux noise 
at $(n_g,f_e)=(0.5,0)$ and found that they are much longer 
than the corresponding characteristic times for the $1/f$ charge noise. 
Thus, we conclude that 
the $1/f$ flux noise plays the least dominant role at the degeneracy point 
for the qubit in the charge-flux regime with $E_J=E_c$.
Moreover, for both $1/f$ charge and flux noises, we found that, 
in the vicinity of the degeneracy point $(n_g,f_e)=(0.5,0)$,
$T_1$, $T_{L0}$, and $T_{L1}$ depend weakly on $n_g$ and $f_e$.
The decoherence time $T_2$ also depends weakly on $f_e$ ($n_g$) 
for the $1/f$ charge (flux) noise, but very strongly on $n_g$ ($f_e$);  
slightly away from the degeneracy point along $n_g$ ($f_e$), the decoherence 
time $T_2$ decreases several orders of magnitudes. 

For clarity, the relaxation and decoherence times $T_1$ and $T_2$ are replotted 
in the inset of Fig.~4 for the $1/f$ charge noise. 
At $\alpha=20$, $T_2\approx 0.5$~$\mu$s, and $T_1\approx 1.6$~$\mu$s. 
Note that this agrees with the experimental results~\cite{VION} 
of the charge-flux qubit with a large left JJ.
Also, it can be seen that the relaxation time remains at $T_1\sim 1.5$~$\mu$s 
until $\alpha\sim 1.3$, while the decoherence time $T_2$ first increases with 
decreasing $\alpha$, then remains at $T_2\sim 3$~$\mu$s (the longest decoherence 
time) when $1.3\alt\alpha\alt 16$, 
and finally falls down for $\alpha\alt 1.3$. 
Therefore, one can optimize the charge-flux qubit in the region 
$1.3\alt\alpha\alt 16$, so that the qubit has the lowest decoherence. 

%

\begin{figure}
\includegraphics[width=3.4in,
bbllx=58,bblly=185,bburx=503,bbury=707]{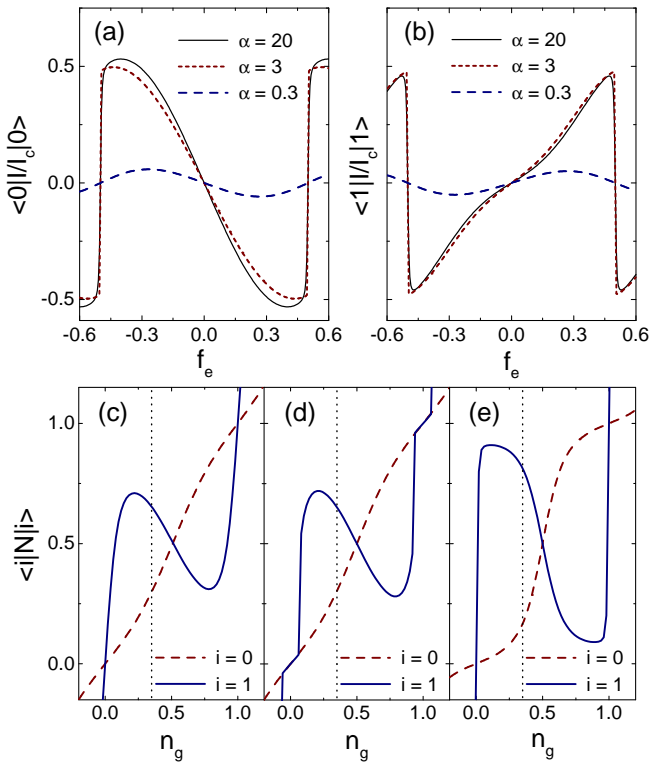}
\caption{(Color online) Circulating currents in the qubit loop versus $f_e$ 
for eigenstates $|i\rangle$, (a) $i=0$ and (b) 1, where $n_g=0.5$. 
The number of Cooper pairs on the island versus $n_g$ at eigenstates $|i\rangle$ 
for (c) $\alpha=20$, (d) 3, and (e) 0.3.}
\label{fig5}
\end{figure}

\section{Efficient quantum measurement}

Finally, we focus on how to raise the readout efficiency. 
Figures 5(a) and 5(b) display the circulating current $I$ in the qubit loop 
at eigenstates $|0\rangle$ and $|1\rangle$. It is clear that the currents
$\langle i|I/I_c|i\rangle$, for $\alpha=20$ and 3, are close to each other. 
This further indicates that the left JJ with $\alpha=3$ 
still behaves like a large JJ barely affecting the CPB qubit. 
Here we consider the readout scheme in Ref.~\onlinecite{VION}, 
where a current pulse is applied to the qubit circuit via
a current source connected in parallel with the left JJ.
This gives rise to an effective capacitance $C_3$ with a larger value 
of $\beta$. 
Thus, the effect of the left JJ on the CPB qubit is further 
weakened because the interaction Hamiltonian $H_I$ 
decreases when increasing $\beta$.
%

For a single left JJ without the right CPB in Fig.~1(a), 
when biased by a current pulse, it switches at 
\begin{equation}
I_{\rm sw}(\alpha)\sim I_{c3}(\alpha),
\end{equation}
with a narrow switching-probability distribution,
from the zero-voltage state to the dissipative nonzero-voltage state. 
Here 
$I_{c3}(\alpha)=\alpha I_c$
is the critical current of the left JJ. However,
when the current pulse is biased to the qubit circuit, i.e., the left JJ 
plus the right CPB [see Fig.1(a)], the left JJ switches 
at 
\begin{equation}
I_{\rm sw}^{(\rm qubit)} = I_{\rm sw}(\alpha)+\langle i|I|i\rangle 
\end{equation}
with probabilities $p_i\,(s_i)$ ($i=0,1$)
which depend on (see, e.g.,~Ref.~\onlinecite{COTT})
\begin{equation}
s_i={I_{\rm sw}+\langle i|I|i\rangle\over I_{c3}} \,.
\end{equation}
In Ref.~\onlinecite{VION}, $\alpha=20$, and the switching-probability difference 
is found to be as small as 
\begin{equation}
|p_0-p_1|\sim 0.1 
\end{equation}
because
\begin{equation}
|s_0-s_1|={|\langle 0|I/I_c|0\rangle-\langle 1|I/I_c|1\rangle|\over\alpha}
\end{equation} 
is small for $\alpha=20$. 
However, when the left JJ becomes smaller, to $\alpha=3$, 
\[
|\langle 0|I/I_c|0\rangle-\langle 1|I/I_c|1\rangle|
\]
remains nearly unchanged, 
but $|s_0-s_1|$ is enlarged about seven times. 
This greatly increases $|p_0-p_1|$ and thus efficiently discriminates the states 
$|0\rangle$ and $|1\rangle$. 

In Figs. 5(c)-5(e), we show the number of Cooper pairs on the island, $\langle i|N|i\rangle$, 
at eigenstates $|i\rangle$, $i=0,1$. For a given $i$, $\langle i|N|i\rangle$
at $\alpha=20$ and 3 are similar to each other but much different from that at $\alpha=0.3$. 
For instance, when $n_g=0.34$ (indicated by a vertical black dotted line), 
the number difference 
\[
\Delta N\,\equiv\, |\langle 0|N|0\rangle-\langle 1|N|1\rangle|
\]
is $\Delta N \sim 0.37$ for both $\alpha=20$ and 3, but increases to 
$\Delta N \sim 0.67$ when $\alpha=0.3$.
Therefore, the readout efficiency for discriminating the states $|0\rangle$ 
and $|1\rangle$ can be much increased at $\alpha=0.3$, 
when a single-electron transistor~\cite{DEVO} is capacitively connected to the island
and used for measuring the quantum states. 
Also, one can effectively couple two CPB qubits with $\alpha=0.3$
by taking advantage of the charge degree of freedom, such as connecting the islands 
in the two qubits via a mutual capacitance (see, e.g., Ref.~\onlinecite{AMIN}). 
This capacitive coupling can be used to reduce decoherence in a logical qubit 
composed of two CPB qubits.~\cite{deco}

\section{Conclusion}

In conclusion, we have studied a hybridized charge-flux qubit 
in which an additional JJ is added into the SQUID loop of the CPB.
The goal is to find a low-decoherence superconducting qubit. 
This is one of the most important open issues since quantum computing 
is possible only if a qubit with long enough decoherence time becomes 
available. 

Currently, $1/f$ noise is believed to be the main source of decoherence 
in a superconducting qubit.
Here we consider the effects of both charge and flux $1/f$ noises 
on the hybridized charge-flux qubit.
We find that the qubit is optimized in the region 
$1.3 \alt \alpha \alt 16$, so that the qubit has the lowest decoherence. 
These results indicate how to optimize a qubit that is expected to have a longer 
decoherence time. 
Moreover, we find that the readout scheme via measuring 
currents, like that in Ref.~\onlinecite{VION}, 
can also be optimized, so that the efficiency for discriminating 
qubit states is much increased.
Furthermore, we show that an efficient readout scheme by measuring charges 
can be achieved as well.             

Note that our studies on the $1/f$ noise use 
the harmonic bath model with a $1/\omega$ spectral density. 
This is valid when the $1/f$ noise is not dominated 
by a few fluctuators strongly coupled to the qubit.
Here, due to the lack of available data for a charge-flux qubit,
the numerical values of $\alpha_q$ and $\alpha_{\Phi}$ in the power spectra of 
$1/f$ charge and flux noises are chosen from the experimental data of 
the charge and flux qubits. Also, for simplicity,
the same frequency cutoff is used for both charge and flux $1/f$ noises. 
In some cases, this might considerably deviate from the realistic samples. 
Thus, more experimental data are needed for giving
a quantitative comparison with realistic samples.

\begin{acknowledgments}
We thank Y. Nakamura, Yu. Pashkin, O. Astafiev, T. Yamamoto, S.Y. Han, 
and Y.X. Liu for discussions.
This work was supported in part by the National Security Agency (NSA) 
and Advanced Research and Development Activity (ARDA) under Air Force 
Office of Research (AFOSR) contract number~F49620-02-1-0334, 
and by the National Science Foundation grant No.~EIA-0130383.
J.Q.Y. was also supported by the 
National Natural Science Foundation of China
grant Nos.~10474013 and 10534060.
\end{acknowledgments}



\begin{references}
%
\bibitem{NAKA} Y. Nakamura, Yu. A. Pashkin, and J.S. Tsai, Nature (London) {\bf 398},
786 (1999); Yu. A. Pashkin, T. Yamamoto, O. Astafiev, Y. Nakamura, and J.S. Tsai, 
{\it ibid.} {\bf 421}, 823 (2003).
\bibitem{HAN} Y. Yu, S.Y. Han, X. Chu, S.I. Chu, and Z. Wang, Science {\bf 296}, 
889 (2002); J.M. Martinis, S. Nam, J. Aumentado, and C. Urbina, 
Phys. Rev. Lett. {\bf 89}, 117901 (2002).
\bibitem{CHIO} I. Chiorescu, Y. Nakamura, C.J.P.M. Harmans, and J.E. Mooij, 
Science {\bf 299}, 1869 (2003).
\bibitem{VION} D. Vion, A. Aassime, A. Cottet, P. Joyez, H. Pothier, C. Urbina,
D. Esteve, and M.H. Devoret, Science {\bf 296}, 886 (2002).
\bibitem{FN} J.Q. You and F. Nori, Phys. Rev. B {\bf 68}, 064509 (2003); 
J.Q. You, J.S. Tsai, and F. Nori, {\it ibid.} {\bf 68}, 024510 (2003);
Y.X. Liu, L.F. Wei, and F. Nori, Europhys. Lett. {\bf 67}, 941 (2004).
\bibitem{Yale} A. Blais, R.S. Huang, A. Wallraff, S.M. Girvin, and R.J. Schoelkopf,
Phys. Rev. A {\bf 69}, 062320 (2004); 
A. Wallraff, D. I. Schuster, A. Blais, L. Frunzio, R.S. Huang, J. Majer, S. Kumar, 
S. M. Girvin, and R. J. Schoelkopf, Nature (London) {\bf 431}, 162 (2004).
\bibitem{FLUX} I. Chiorescu, P. Bertet, K. Semba, Y. Nakamura,
C.J.P.M. Harmans, and J.E. Mooij, Nature (London) {\bf 431}, 159 (2004).
\bibitem{ORL} T.P. Orlando, J.E. Mooij, L. Tian, C.H. van der Wal, L.S. Levitov,
S. Lloyd, and J.J. Mazo, Phys. Rev. B {\bf 60}, 15398 (1999).
\bibitem{YOU} J.Q. You, Y. Nakamura, and F. Nori, Phys. Rev. B {\bf 71},
024532 (2005); Y.X. Liu, J.Q. You, L.F. Wei, C.P. Sun, and F. Nori, 
Phys. Rev. Lett. {\bf 95}, 087001 (2005).
\bibitem{BURK} G. Burkard, R.H. Koch, and D.P. DiVincenzo, Phys. Rev. B
{\bf 69}, 064503 (2004).
\bibitem{BERT} P. Bertet, I. Chiorescu, G. Burkard, K. Semba, C.J.P.M. Harmans, 
D.P. DiVincenzo, and J.E. Mooij, Phys. Rev. Lett. {\bf 95}, 257002 (2005).
\bibitem{PAL} E. Paladino, L. Faoro, G. Falci, and R. Fazio, Phys. Rev. Lett. 
{\bf 88}, 228304 (2002).
\bibitem{GAS}
Yu.M. Galperin, B.L. Altshuler, and D.V. Shantsev, cond-mat/0312490;
L. Faoro, J. Bergli, B.L. Altshuler, and Y.M. Galperin, Phys. Rev. Lett. 
{\bf 95}, 046805 (2005).
\bibitem{NG} A.K. Nguyen and S.M. Girvin, Phys. Rev. Lett. {\bf 87}, 
127205 (2001).
\bibitem{MS} Yu. Makhlin and A. Shnirman, JETP Lett. {\bf 78}, 497 (2003);
Phys. Rev. Lett. {\bf 92}, 178301 (2004).
\bibitem{NAKA1} Y. Nakamura, Yu.A. Pashkin, T. Yamamoto, and J.S. Tsai, 
Phys. Rev. Lett. {\bf 88}, 047901 (2002).
\bibitem{COTT} A. Cottet, D. Vion, A. Aassime, P. Joyez, D. Esteve, and M.H. Devoret,
Physica C {\bf 367}, 197 (2002).
\bibitem{DEVO} M.H. Devoret and R.J. Schoelkopf, Nature (London) 
{\bf 406}, 1039 (2000).
\bibitem{AMIN} M.H.S. Amin, Phys. Rev. B {\bf 71}, 024504 (2005).
\bibitem{deco} J.Q. You, X. Hu, and F. Nori, Phys. Rev. B {\bf 72}, 144529 (2005).


%
\end{references}
\end{document}